\documentclass[article]{svmult}
\usepackage{amssymb}
\usepackage{amsmath}
\usepackage{mathrsfs}

\usepackage{epsfig}
\usepackage{graphics}
\usepackage{float}
\usepackage{tabularx}
\usepackage{color}
\newcolumntype{C}[1]{>{\centering\arraybackslash}p{#1}}
\begin{document}
\title*{Primordial Generation of Magnetic Fields}
\author{Jitesh R. Bhatt, Arun Kumar Pandey}
\institute{Jitesh R. Bhatt \at Theoretical Physics Division, Physical Research Laboratory, Ahmedabad 380009, India, \email{jeet@prl.res.in} \and
Arun Kumar Pandey \at Theoretical Physics Division, Physical Research Laboratory, Ahmedabad 380009, India \& 
Department of Physics, Indian Institute of Technology, Gandhinagar, Ahmedabad, India, \email{arunp@prl.res.in}}
\maketitle
\def\be{\begin{equation}}
\def\ee{\end{equation}}
\def\al{\alpha}
\def\bea{\begin{eqnarray}}
\def\eea{\end{eqnarray}}
\def\beas{\begin{eqnarray*}}
\def\eeas{\end{eqnarray*}}
\abstract{
 We reexamine generation of the primordial magnetic fields, at temperature $T>80$TeV, by applying 
 a consistent kinetic theory framework which is suitably modified to take the quantum anomaly into account.
 The modified kinetic equation can reproduce the known quantum field theoretic results upto the leading
 orders. We show that our results qualitatively matches with the earlier results obtained using heuristic
 arguments.  The modified kinetic theory can give the instabilities responsible
 for generation of the magnetic field due to chiral imbalance in two distinct regimes: a) when the collisions
 play a dominant role and b) when the primordial plasma can be regarded as collisionless. We argue
 that the instability developing in the collisional regime can dominate over the instability 
 in the collisionless regime.}
\title*{Primordial Generation of Magnetic Fields}
\author{Jitesh R. Bhatt, Arun Kumar Pandey}
\institute{Jitesh R. Bhatt \at Theoretical Physics Division, Physical Research Laboratory, Ahmedabad 380009, India, \email{jeet@prl.res.in} \and
Arun Kumar Pandey \at Theoretical Physics Division, Physical Research Laboratory, Ahmedabad 380009, India \& 
Department of Physics, Indian Institute of Technology, Gandhinagar, Ahmedabad, India, \email{arunp@prl.res.in}
}
\maketitle
\section{Introduction} 
 
  Observations suggest that we live in a magnetized Universe: Magnetic fields are present from stellar scales to
  the intergalactic scales. 
  However, it is yet not clear that how these magnetic field arises. One of the possible thought is that 
  their origin is due to some process in the 
early universe like inflation\cite{Carroll:1989vb} or phase transitions\cite{Vachaspati:1991nm} etc. It may also possible 
that the magnetic fields may not have any primordial origin but they may be generated during a gravitational 
collapse\cite{Kulsrud:1996km}. 
Currently the generation and dynamics of the primordial magnetic field is one of the most intriguing problem in cosmology
\cite{Kandus:2010nw}.

	In recent time, there has been great interests in generation of primordial fields by quantum anomaly\cite{Tashiro:2012mf}.
A chiral imbalance of the leptons can occur in the very early universe, due to some electroweak(EW) anomalous processes like parity violating
decays of massive particles, just before EW scale\cite{Cornwall:1997ms}. It has been shown that, the presence of chiral imbalance creates 
instability in the the hot matter\cite{Akamatsu:2013pjd}. 
At temperatures $T>$ 80~TeV($\sim T_R$), the chirality flipping 
processes are highly suppressed and hence  asymmetry between the right-handed and
left-handed particle is preserved. In the regime $T>T_{R}$, the EW symmetry is unbroken and the generated field will be $U(1)$ 
hypermagnetic(hypercharge) fields\cite{Vachaspati:2001nb}. After EW phase transition, these hypermagnetic field will be converted into the electromagnetic field. 

	Recently there has been growing interests in including the parity violating effects into a kinetic theory formalism. It was 
found that the chiral anomaly can be incorporated in a kinetic theory framework by including the Berry curvature correction\cite{Son:2012wh}.
The resulting theory can reproduce the triangle anomaly and provide the descriptions  
for the chiral magnetic and chiral vorticity effects\cite{Son:2012zy}.
In this work we apply the Berry curvature modified kinetic theory to the chiral plasma for the study of 
generation of primordial magnetic field. In this work, we discuss the generation and evolution of
magnetic field in two regions: a) where collision play a dominant role
i.e. $\nu_c>>\omega>>k$ where $\nu_c$ is the collision frequency, $\omega$ and $k$ respectively denote
the typical frequency and wave-number for the perturbations.
b)   And in the collisionless regime $k>>\omega>>\nu_c$. 
The instabilities can grow in these two regimes at the expense of the chiral-imbalance and surprising the
maximum growth rates for both regimes occur at the similar length scales.
In this work, we analyze the characteristics of these instabilities within the framework of the modified kinetic
theory and discuss the relationship between  our results and the earlier results.
The manuscript is organized into two sections. In first section we discussed the generation of primordial magnetic field
incorporating Berry correction. Second section contains results and discussion.
\section{Generation of the primordial magnetic field}
In this work we solve a coupled system of the Maxwell equations and the modified kinetic theory  for relativistic particles 
in the spatially flat Friedmann-Lema\^{\i}tre-Robertson-Walker universe. Conformal flatness of this space-time insures that equations in the 
conformal space have same forms as those of the flat space-time\cite{Holcomb:1989tf}. A conformal metric can be written as:
\begin{equation}
 ds^{2}=a^2(\eta)(- d\eta^2+ dx^2 +dy^2 +dz^2).
\end{equation}
where conformal time $\eta$ is related with proper time $t$ by $\eta =\int dt/a^2(t)$ The quantities like (hyper)-electric $\boldsymbol{E}$, 
(hyper)-magnetic $\boldsymbol{B}$ and the current density $\boldsymbol{J}$ measured by the comoving observer are related with quantities 
in the conformally flat space quantities 
through scale factor as: $\boldsymbol{E}= a^2\boldsymbol{\mathcal{E}}$, $\boldsymbol{B}= a^2\boldsymbol{\mathcal{B}}$ and 
$\boldsymbol{J}= a^2\boldsymbol{\mathcal{J}}$.

By inclusion of Berry correction to the kinetic theory, the distribution function as well as kinetic equations modifies. The modified kinetic equation under relaxation
time approximation, at the first order in perturbation is:
\begin{equation} \label{kinetic equation}
 \bigg(\frac{\partial }{\partial\eta}+{\boldsymbol{v}}\cdot\frac{\partial}{\partial \boldsymbol{r}}\bigg)f_i+
 \bigg(e^{i}\boldsymbol{E}+e^{i}(\boldsymbol{v}\times \boldsymbol{B})- 
 \frac{\partial{\epsilon^{i}_p}}{\partial{\boldsymbol{r}}}\bigg)\cdot\frac{\partial f_i}{\partial \boldsymbol{p}}=
 \bigg(\frac{\partial f_i}{\partial\eta}\bigg)_{coll.}, 
\end{equation}
The index $i$, in above equation represents different species of the leptons. $\boldsymbol{\Omega_{p}}$ comes for Berry curvature and defined by 
$\boldsymbol{\Omega_{p}}= \pm \boldsymbol{p}/(2p^3)$. Positive and negative signs comes for right handed fermions and for left handed fermions respectively. 
$\epsilon_{\boldsymbol{p}}$ is defined as  $\epsilon_{\boldsymbol{p}}^i=p(1-e^i\boldsymbol{B}\cdot\boldsymbol{\Omega_{p}}^i)$ with $p=|\boldsymbol{p}|$. 
Depending on the species charge $e$, energy of the particles $\epsilon_p$, Berry curvature $\boldsymbol{\Omega_p}$ and form of the distribution function
$f$ changes. In the absence of Berry correction i.e. $\boldsymbol{\Omega_p}=0$, 
above equation  reduces to the Vlasov equation when the collision term on the right hand side of 
equation(\ref{kinetic equation}) is absent. With inclusion of Berry curvature, current density also modifies as:
\begin{align} \label{current1}
 \boldsymbol{J}_{i}= -e^i\int\frac{d^3p}{(2\pi)^3}\bigg[\epsilon_{\boldsymbol{p}}^i\frac{\partial {f}_{i}}{\partial \boldsymbol{p}}
 +e^i(\boldsymbol{\Omega_{\boldsymbol{p}}^i}\cdot\frac{\partial{f}_{i}}{\partial{\boldsymbol{p}}}) \epsilon_{\boldsymbol{p}}^i \boldsymbol{B}+
 \epsilon_{\boldsymbol{p}}\boldsymbol{\Omega_{\boldsymbol{p}}^i}\times\frac{\partial{f}_{i}}{\partial \boldsymbol{r}}\bigg]
\end{align}
Above temperature $T~>~T_R$, the chiral plasma remains in thermal equilibrium. And masses of the plasma particles can be ignored. So 
plasma in equilibrium can be considered to be in homogeneous and isotropic state. So distribution function for different species in thermal equilibrium
can be written as $f_{0i}(p)=[exp(\frac{p-\mu_i}{T})+1]^{-1}$. Now let us suppose that, $\delta f_i$ is
fluctuations in the distribution functions of the particle species around it's equilibrium distribution, then 
perturbed distribution function can be written as: $f_{i}({\boldsymbol{r, p}}, \eta)= f_{0i}(p)+ \delta f_{i}({\boldsymbol{r, p}}, \eta)$.
Now one can obtain the total current $\boldsymbol{J}$ by adding the contribution from all the species of the particle $\boldsymbol{J}_a$. 
Total current comes out:
\begin{eqnarray} \label{j total}
 \boldsymbol{J}_{\boldsymbol{k}}= &-m_D^2 \int \frac{d\Omega}{4\pi} \frac{\boldsymbol{v}(\boldsymbol{v}.
 {\boldsymbol{E_{k}}})}{i(\boldsymbol{k.v}-\omega-i \nu_c)}
 -\frac{c_D^2}{2}\int\frac{d \Omega}{4\pi} \bigg(\frac{\boldsymbol{v} (\boldsymbol{v.B_{k}})(\boldsymbol{k.v})-
 (\boldsymbol{v\times k})(\boldsymbol{v.E_{k}})}{(\boldsymbol{k.v}-\omega-i \nu_c)}+ \boldsymbol{B}_{k}\bigg) \nonumber \\
 &-\frac{i g_D^2}{4}\int \frac{d\Omega}{4\pi} \frac{(\boldsymbol{v \times k})(\boldsymbol{v.B_k})(\boldsymbol{k.v})}{(\boldsymbol{k.v}-\omega-i \nu_c)}
 -\frac{h_D^2}{2} \int \frac{d\Omega}{4\pi}\{\boldsymbol{B}_k-\boldsymbol{v}(\boldsymbol{v. B_k})\}
\end{eqnarray}
To get the expression (\ref{j total}) from eq.(\ref{current1}), we solved eq.(\ref{kinetic equation}) to get distribution function.
Here $\Omega$ represent angular integrals. In eq.(\ref{j total}), we have defined $m_D^2 =e^2\int \frac{p^2 dp}{2\pi^2}\frac{df_0}{dp}$, 
$c_D^2 =e^2\int \frac{p dp}{2\pi^2}\frac{df_0}{dp}$, $g^2_D=e^2\int \frac{dp}{2\pi^2}\frac{df_0}{dp}$ \& $h^2_D=e^2\int \frac{dp}{2\pi^2}f_0$
and $f_0= \sum_{a} f_{oi}$.
Plasma with chirality imbalance are known to have instabilities that can generate magnetic fields 
in two different regimes: (i) for the case when $k<<\omega<<\nu_c$ \cite{Joyce:1997uy} and
(ii) in the quasi-static limit i.e. $\omega<<k$ and $\nu_c=0$ \cite{Akamatsu:2013pjd}.
In this section we analyze how the magnetic fields evolve in the plasma due to these instabilities, 
within the modified kinetic theory frame work.  
Using Maxwell's equation and current expression given above, one can get diffusivity equation for chiral plasma in the regime $k<<\omega<<\nu_c$ as:
\begin{equation} \label{B_kw vector}
 \frac{\partial \boldsymbol{B}_k}{\partial \eta}+\left(\frac{3\nu_c}{4\pi m_d^2}\right)k^2\boldsymbol{B}_k 
 -i\left(\frac{\alpha\Delta\mu}{\pi m_D^2}\right)\left(\boldsymbol{k}\times\frac{\partial \boldsymbol{B}_k}{\partial \eta}\right)
+i\frac{4\alpha \nu_c \Delta\mu}{\pi m_D^2} (\boldsymbol{k\times B_k})=0.
\end{equation}
We can solve this equation without a loss of generality by considering the propagation vector $\boldsymbol{k}$ in $z-$direction and
the magnetic field having components perpendicular to $z-$ axis. After defining two new variables: ${B}_k= (B_k^1+ i B_k^2)$ and 
$\tilde{B^{\prime}_k}= (B_k^1- i B_k^2)$ one can rewrite eq.(\ref{B_kw vector}) as:
\begin{eqnarray}
 \frac{\partial\tilde{B}_{k}}{\partial \eta}+\frac{3\nu_c}{4\pi m_d^2}k\left[\frac{k-\frac{16\alpha \Delta\mu}{3}}
 {(1+\frac{\alpha \Delta \mu k}{\pi m_D^2})}\right]\tilde{B}_{k}=0, \label{mode1} \\
  \frac{\partial\tilde{B}_{k}^{\prime}}{\partial \eta}+\frac{3\nu_c}{4\pi m_d^2}k\left[\frac{k+\frac{16\alpha \Delta\mu}{3}}
 {(1-\frac{\alpha \Delta \mu k}{\pi m_D^2})}\right]\tilde{B}_{k}^{\prime}=0.\label{mode2}
\end{eqnarray}
 Another regime where chiral imbalance instability can occur is in the quasi-static limit when
$\omega << k$ and $\nu_c=0$ \cite{Akamatsu:2013pjd}. In this case the diffusivity equation comes out:
\begin{equation}
 \frac{\partial \boldsymbol{B_k}}{\partial \eta}+ \frac{k^2}{4\pi \sigma_1}\boldsymbol{B_k}- i
 \frac{\alpha T \delta}{\pi \sigma_1}\left(\boldsymbol{k}\times\boldsymbol{B_k}\right)=0
\end{equation}
where $\sigma_1 =\pi m_D^2/2k$. One can get decoupled equation as in the case of $\nu_c>\omega>k$. Which is:
\begin{eqnarray}
 \frac{\partial\tilde{B}_{k}}{\partial \eta}+\Bigg[\frac{k^2-\frac{4\alpha \Delta \mu k}{3}}{\frac{\pi m_D^2}{2k}}
 \Bigg]\tilde{B}_{k}=0, \label{mode11}\\
 \frac{\partial\tilde{B}_{k}^{\prime}}{\partial \eta}+\Bigg[\frac{k^2+\frac{4\alpha \Delta \mu k}{3}}{\frac{\pi m_D^2}{2k}}
 \Bigg]\tilde{B_k}^{\prime}=0.\label{mode22} 
\end{eqnarray}
\section{Results and discussion}
 Equations (6-7, 9-10) describe the evolution of the primordial magnetic field due to quantum anomaly for both the collisional
 and collisionless regimes. For the case when $\nu_c>>\omega>>k$, modes described by eq.(\ref{mode1}), 
becomes unstable when $k<(16/3)\alpha\Delta\mu$. The instability has the maximum growth rate of 
$\gamma_1\sim \frac{16}{3\pi}\frac{T^2 \delta^2}{m^2_D}\nu_c$
for $k^1_{max}\sim \frac{8\alpha T \delta}{3}$. 
Eq.(\ref{mode2}) can also give the instability if the denominator of the second term on the right hand side become
negative. But this possibility is ruled out for the  present case.
For $\tilde{B}^{\prime}_k$ in eq.(\ref{mode2}), the modes are damped if 
the condition $\left(\frac{\alpha T \delta }{\pi m^2_D}\right)k<1$ is satisfied. 
This condition is always satisfied since $\alpha<<1$ , $k/m_D<1$ and $T/m_D\sim O(1)$. Thus eq. (\ref{mode2}) can not
give the unstable modes. The value of $k^1_{max}$ is of same order of magnitude as the one given in Ref.\cite{Tashiro:2012mf}. 
But in the collisionless regime, the modes  of $\tilde{B}_k$, as seen from eq.(\ref{mode11}), 
become unstable for $k< 4\alpha \Delta \mu/ 3$.
The instability has maximum growth rate $\gamma_2\sim \frac{\alpha^3}{2\pi}\left( \frac{T\delta}{m_D}\right)^2(T\delta)$ at 
$k^2_{max}\sim \frac{8\alpha T\delta}{9}$. Eq.(\ref{mode22}) describes the purely damping modes.
It has been found that, instability in the two regime has almost same length scale for which their modes
start growing. So it is interesting to ask, if instabilities in the two regime occurs almost at the same length scale, 
then which one will win.
For this we compare the maximum growing rates in the two regimes and we found that
$\frac{\gamma_1}{\gamma_2}\sim 10(\alpha \delta)^{-1}$. Clearly for $\delta<<1$, it is the collisional plasma whose modes will grow 
faster than the modes of the collisionless plasma. However for the chiral plasma, where $\delta>>1$ the situation may be reversed.
The magnetic field will continue to grow at the cost of the chiral charge. This can be seen from the anomaly equation which at $T>80$~TeV 
gives $ n_L-n_R+2\alpha\mathcal{H}$=constant, where, $n_{L,R}=\frac{\mu_{L,R}T^2}{6}$ and $\mathcal{H}$ is the magnetic helicity.
   
  In conclusion, we have applied the kinetic theory with the Berry curvature correction
to study origin of the primordial magnetic field due to anomaly.
We have incorporated the effect of collisions using the relaxation time approximation. Further we have shown that the instability that
can be present in a collisionless chiral plasma may not grow as fast as one found
in presence of collision (in Ref.\cite{Joyce:1997uy}) when $\delta<<1$.



\end{document}